\documentclass[a4paper,12pt]{article}
\usepackage{graphicx,epsfig}
\usepackage[active]{srcltx}
\def\dr{Rindler-Rindler }
\def\bg{Bogoliubov }

\title{Quantum field theory in the Rindler-Rindler spacetime}

\author{Sanved Kolekar\footnote{sanved@iucaa.ernet.in} ~ and T.~Padmanabhan\footnote{paddy@iucaa.ernet.in}\\
IUCAA, Pune University Campus, Ganeshkhind,\\
 Pune 411007, INDIA. \\ 
}

\date{\today}
\begin{document}
\maketitle
\begin{abstract}
It is well known that Minkowski vacuum appears as a thermal bath in the Rindler spacetime when the modes on the left wedge are traced out. We introduce the concept of a Rindler-Rindler spacetime, obtained by a further coordinate transformation from the Rindler spacetime, in a manner similar to  the transformation from inertial to Rindler frame. We show that the Rindler vacuum appears as a thermal state in the Rindler-Rindler frame.  Further, the spectrum of particles seen by the Rindler-Rindler observers in the original Minkowski vacuum state is shown to be identical to that seen by detector  accelerating through a real thermal bath. Thus the Davies-Unruh effect acts as a proxy for a real thermal bath, for a certain class of observers in the Rindler-Rindler spacetime. We interpret this similarity as indicating further evidence of the indistinguishablity between thermal and quantum fluctuations along the lines of the recent work in arXiv:1308.6289. The implications are briefly discussed.

\end{abstract}

\section{Introduction}

It is well-known that for an uniformly accelerated observer, the  fluctuations of a quantum field in the inertial vacuum state appear to be identical to the thermal fluctuations of an real thermal bath indicating some kind of equivalence between thermal and vacuum fluctuations \cite{Davies, Unruh}. There  have also been suggestions  that quantum and statistical fluctuations, including thermal ones, are essentially identical \cite{Smolin} (see also \cite{cordoba}). In a recent paper \cite{san}, we explored this relationship between thermal and quantum fluctuations beyond the context of  the vacuum state. We showed that, when  the uniformly accelerated observer is moving through a genuine thermal bath, he is unable to distinguish between the thermal fluctuations and the fluctuations generated due to the correlations between the quantum field in the two Rindler wedges. In particular, we calculated the reduced density matrix for an uniformly accelerated observer moving with acceleration $a = 2\pi/ \beta$ in a thermal bath of temperature $1/\beta^\prime$ and showed that it is symmetric in the acceleration temperature $1/\beta$ and the thermal bath temperature $1/\beta^\prime$. Thus, we concluded that, purely within the thermodynamic domain, it is not possible to distinguish between the thermal effects due to $\beta$ and those due to $\beta^\prime$.

In the above mentioned analysis, we considered the  thermal fluctuations to be those of a real thermal bath in a flat spacetime. On the other hand, we know that the Minkowski vacuum fluctuations themselves are identical to those of a genuine thermal fluctuations for the uniformly accelerated observer (see also \cite{Takagi}). So the question arises as to whether the similar result will arise if a person is accelerating \textit{with respect to a Rindler frame}; i.e, can we use the Rindler frame (in which Minkowski vacuum appears as a thermal bath) as a proxy for a real thermal bath. 
More precisely, this issue would require us to investigate the following two questions. (i) Is there a trajectory in the right Rindler wedge such that the \textit{Rindler vacuum state} appears to be thermal, at a temperature $1/\beta^\prime$ (corresponding to some parameter characterizing the trajectory), to an observer in an accelerating trajectory in the Rindler frame? (ii) If  such a trajectory exists (and we will find that it does),  how does this  observer accelerating with respect to the Rindler frame perceive the  \textit{Minkowski vacuum state}?. Since Minkowski vacuum appears as a thermal bath in the standard Rindler observer, it would be interesting if one finds that the observer accelerating with respect to the Rindler frame obtains results similar to those found by an observer accelerating through a thermal bath. In particular, it would be nice to see whether, in this case too, a symmetry between $\beta$ and $\beta^\prime$ exists. Such a  symmetry would further strengthen the equivalence between thermal and quantum fluctuations.

Addressing the above two issues is the main focus of this paper. Before proceeding, we mention that issues are not straightforward and one requires to take into account the following points:
For a pure state to appear as thermal to some observer, there must be a tracing operation involved, that is, one must ignore a fraction of the total degrees of freedom of the quantum field to obtain a mixed state starting from a pure state. Hence, the required accelerating trajectory in the right Rindler wedge $ X > | T|$ must be further confined in the Rindler spacetime, say the wedge $x > |t|$, where $x,t$ are the right Rindler co-ordinates.   

Further note that there are only ten linearly independent Killing vectors for the Minkowski spacetime; the Rindler trajectory arises as the integral curve of the boost Killing vector which leads to a stationary Planckian spectrum of Rindler particles in the Minkowski vacuum. Hence, it is not obvious whether there exists some other  timelike Killing vector constructed from the combination of the known Killing vectors whose integral curves will be restricted to the $t > |x|$ wedge \cite{probes}. One is then led to the remaining alternative where the required trajectory has to be a non-stationary or non-uniformly accelerating trajectory. 
 This raises further issues about particle definition etc. due to the non-static nature of the resultant metric.

We attempt to address these issues in this paper. In section 2, we show that there exists an trajectory characterized by a parameter $g^\prime = 2\pi/ \beta^\prime$ in the Rindler spacetime such that the Rindler vacuum appears to be thermally populated at time $T = 0$ with temperature $1/\beta^\prime$. We  calculate the spectrum seen by this observer when the quantum field is in the Minkowski vacuum state. We find that the resulting spectrum is symmetric in $1/\beta^\prime$ and the Davies-Unruh temperature $1/\beta$. In section 3, we show that this spectrum is identical to the spectrum of an uniformly accelerating detector in flat spacetime coupled to an inertial thermal bath. The conclusions are discussed in section 4.  

We will work in $1+1$ dimensions. We use units with $c=k_B = \hbar = 1$.

\section{Rindler-Rindler spacetime and quantum fields}

In this section, we use the \bg co-efficients formalism to investigate the required trajectory such that the Rindler vacuum appears to be thermally populated to the observer at a chosen instant of time $T=0$. 
We will briefly describe the setup needed to find such a trajectory. 

\subsection{Minkowski vacuum from the Rindler perspective}

This is standard result which we will briefly summarize to set up the background formalism. Consider a $(1+1)$ dimensional Minkowski spacetime. Let $X,T$ denote the usual set of Minkowski coordinates. The Minkowski metric is given as  
\begin{equation}
ds^2 = (-dT^2 + dX^2)
\end{equation}                                                                                                                                                                                                                                   
with the the solution of the Klein Gordon field equation given by the usual plane wave mode solutions in this conformal metric with the conformal factor being unity. The quantized field can be written as
\begin{equation}
 \phi(X,T) = \int \frac{dk}{2\pi} \left( a_{(0)k}\frac{e^{i(kX-\omega_kT)}}{\sqrt{2\omega_k}} + h.c. \right)
\end{equation}
Next we perform the usual Rindler transformation in the right Rindler wedge
\begin{eqnarray}
X = \frac{e^{gx_1}}{g}\cosh{gt_1} \; \; \qquad
T = \frac{e^{gx_1}}{g}\sinh{gt_1}
\end{eqnarray}                                                                                                                                                                                                                       
to get the Rindler metric in the following conformal form
\begin{equation}
ds^2 = e^{2gx_1}(-dt_1^2 + dx_1^2)
\end{equation}    
It is well known that in $1+1$ dimensional spacetime, due to the conformal nature of the metric, the scalar field solution can be expressed as plane wave modes in terms of the Rindler conformal  coordinates as 
\begin{equation}
\phi(x_1,t_1) = \int \frac{dp}{2\pi} \left( a_{(1)p}\frac{e^{i(px_1-\omega_pt_1)}}{\sqrt{2\omega_p}} + h.c. \right)
\end{equation}
Further, the positive and negative frequency modes of the Minkowski and Rindler observer mix with each other leading to non-trivial \bg coefficients. They are defined through the following relation as
\begin{equation}
\frac{e^{i(kX-\omega_kT)}}{\sqrt{2\omega_k}}  = \int \frac{dp}{2\pi} \left( \alpha_{(10)}(k,p)\frac{e^{i(px_1-\omega_pt_1)}}{\sqrt{2\omega_p}} + \beta_{(10)}(k,p) \frac{e^{-i(px_1-\omega_pt_1)}}{\sqrt{2\omega_p}} \right)
\label{bgdefmnksrind}
\end{equation}
Using the Klein-Gordon product, these are found out to be \cite{unruhdet}
\begin{enumerate}
 \item For $p>0$
\begin{eqnarray}
\alpha(k,p) &=& \frac{\theta(k)}{2\pi g} \sqrt{\frac{\omega_p}{\omega_k}} \, e^{\frac{\pi \omega_p}{2g}} \left( \frac{g}{\omega_k} \right)^\frac{i\omega_p}{g} \Gamma(\frac{i\omega_p}{g}) \\
\beta(k,p) &=& \frac{-\theta(k)}{2\pi g} \sqrt{\frac{\omega_p}{\omega_k}} \, e^{\frac{-\pi \omega_p}{2g}} \left( \frac{g}{\omega_k} \right)^\frac{-i\omega_p}{g} \Gamma(\frac{-i\omega_p}{g})
\end{eqnarray}

\item For $p<0$
\begin{eqnarray}
\alpha(k,p) &=& \frac{\theta(-k)}{2\pi g} \sqrt{\frac{\omega_p}{\omega_k}} \, e^{\frac{\pi \omega_p}{2g}} \left( \frac{g}{\omega_k} \right)^\frac{-i\omega_p}{g} \Gamma(\frac{-i\omega_p}{g}) \\
\beta(k,p) &=& \frac{-\theta(-k)}{2\pi g} \sqrt{\frac{\omega_p}{\omega_k}} \, e^{\frac{-\pi \omega_p}{2g}} \left( \frac{g}{\omega_k} \right)^\frac{i\omega_p}{g} \Gamma(\frac{i\omega_p}{g})
\label{bgoriginal}
\end{eqnarray}
\end{enumerate}
The expectation value of Rindler number operator $N= a^\dagger_{(1)k}a_{(1)k}$ in the Minkowski vacuum $|0_M \rangle$ leads to the well known Planckian distribution
\begin{eqnarray}
{\cal N} &=& \langle 0_M | a^\dagger_{(1)k}a_{(1)k} |0_M \rangle =\int d k^\prime |\beta_{(10)}({k^\prime, k})|^2 \nonumber \\
&=& \frac{1}{e^{\beta \omega_{{\bf k}}} -1} 
\end{eqnarray}
where $\beta^{-1}$ is the temperature of the Unruh bath. 

\subsection{Rindler vacuum from \dr\ perspective}

The two ingredients needed for the above result are: (i) the solution as plane wave modes, a fact guaranteed by the conformal nature of the metric and (ii) the Rindler-like transformation essentially exponenciates the Rindler frequencies with respect to  the inertial frequencies.
These facts suggest that we can now perform a second  transformation, similar to that in the Rindler case, to arrive at the Rindler-Rindler spacetime \footnote{We call it the \textit{Rindler-Rindler} spacetime based on the Rindler form of the co-ordinate transformations involved twice. Here, one should note that the usual Rindler transformation from the full Minkowski spacetime to the right (or left) Rindler wedge includes a host of features including the time co-ordinate $t_1$ appearing in the transformation being a proper time along a stationary trajectory. The resulting spacetime (the Rindler wedge) is not a geodesic complete spacetime and is only a part of the Minkowski spacetime. Hence, when one performs another Rindler-like transformation in the Rindler wedge itself not all the features associated with the Rindler transformation (in the full Minkowski spacetime) follow, particularly, the time co-ordinate $t_2$ being also related to a timelike Killing vector.}. We take: 
\begin{eqnarray}
x_1 &=& \frac{e^{g^\prime x_2}}{}g^\prime\cosh{g^\prime t_2} \\
t_1 &=& \frac{e^{g^\prime x_2}}{g^\prime}\sinh{g^\prime t_2}
\end{eqnarray}                                                                                                                                                                                                                       
in the $t> |x|$ wedge region of the Rindler spacetime. The metric then becomes
\begin{equation}
ds^2 = e^{2gx_1}e^{2g^\prime x_2}(-dt_2^2 + dx_2^2)
\label{doublerindmetric}
\end{equation}    
which we shall refer to as the Rindler-Rindler spacetime. Again, due of the conformal nature of the above metric, the  solutions to the field equations in the right Rindler-Rindler wedge are plane wave modes
\begin{equation}
\phi(x_2,t_2) = \int \frac{dm}{2\pi} \left( a_{(2)m}\frac{e^{i(mx_2-\omega_mt_2)}}{\sqrt{2\omega_m}} + h.c. \right)
\end{equation}
Thus, it is easy to check that the two criteria mentioned above is satisfied and hence the \bg coefficients defined through the relation
\begin{equation}
\frac{e^{i(px_1-\omega_pt_1)}}{\sqrt{2\omega_p}} = \int \frac{dm}{2\pi} \left( \alpha_{(21)}(p,m)\frac{e^{i(mx_2-\omega_mt_2)}}{\sqrt{2\omega_m}} + \beta_{(21)}(p,m) \frac{e^{-i(mx_2-\omega_mt_2)}}{\sqrt{2\omega_m}} \right)
\label{bgdefrinddbrind}
\end{equation}
are same as the \bg coefficients given in Eq.[\ref{bgoriginal}]. The expectation value of of the number operator $N= a^\dagger_{(2)k}a_{(2)k}$ in the Rindler vacuum $|0_R \rangle$ is equal to the 
\begin{eqnarray}
{\cal N}_{21} &=& \langle 0_R | a^\dagger_{(2)k}a_{(2)k} |0_R \rangle =\int d k^\prime |\beta_{(21)}({k^\prime, k})|^2 \nonumber \\
&=& \frac{1}{e^{\beta^\prime \omega_{{\bf k}}} -1} 
\end{eqnarray} 
We now have the result that the  \textit{Rindler vacuum}   appears to be thermal to the Rindler-Rindler observer evaluated at the $t=0$ and hence $T=0$ hyper-surface. In-fact, if one continues to perform such Rindler like transformations on the corresponding right wedges $n$ times, each time halving the spacetime, then one has a chain of Rindler-Rindler-Rindler ....($n$ times)
spacetimes with  the vacuum state of each appearing as a thermal bath for the next case.\\

However, one must note that the time co-ordinate $t_2$ with respect to which the positive and negative frequency  modes were defined in the \dr\ spacetime  does not correspond to the proper time of $x_2 =$ constant observers. This is in contrast with the case of the  Rindler frame where the co-ordinate time $t_1$ is the proper time of the Rindler observer. The reason for this difference being that the \dr metric in Eq.[\ref{doublerindmetric}] is not static. However, we \textit{can} find a trajectory such that the proper time $\tau$ along the trajectory corresponds to the time co-ordinate $t_2$. This can be done as follows. Let $x_2(t_2) = x_2(\tau) = y(\tau)$ denote such a trajectory. We know that it must satisfy the normalization relation  $u^i (\tau) u_i(\tau) = -1$. Using Eq.[\ref{doublerindmetric}], it is easy to check that we get the following first order non-linear differential equation
\begin{eqnarray}
-1 = e^{2(g/g^\prime) e^{g^\prime y} \cosh{g^\prime \tau}} e^{2 g^\prime y} \left( -1 + ({\dot y})^2 \right)
\end{eqnarray}
Rearranging, we can express the above equation as
\begin{eqnarray}
{\dot y}^2  = 1 - e^{-2(g/g^\prime) e^{g^\prime y} \cosh{g^\prime \tau}} e^{-2 g^\prime y}
\label{diffeqn}
\end{eqnarray}
It turns out that the above differential equation does not have any known analytic solutions (as far as we know) but the solution does exist proving the existence of such an observer. We hope to study this trajectory in detail in a future work.

\subsection{Minkowski vacuum from \dr\ perspective}

We now proceed to calculate the expectation value of occupation operator $N= a^\dagger_{(2)k}a_{(2)k}$ in the \textit{Minkowski vacuum} $|0_M \rangle$. Let us define the \bg coefficients between the Minkowski and \dr\ modes through the relation
\begin{equation}
\frac{e^{i(kX-\omega_kT)}}{\sqrt{2\omega_k}}  = \int \frac{dm}{2\pi} \left( \alpha_{(20)}(k,m)\frac{e^{i(mx_2-\omega_mt_2)}}{\sqrt{2\omega_m}} + \beta_{(20)}(k,m) \frac{e^{-i(mx_2-\omega_mt_2)}}{\sqrt{2\omega_m}} \right)
\label{bgdefmnksdbrind}
\end{equation}
Using Eqn.[\ref{bgdefrinddbrind}] in Eqn.[\ref{bgdefmnksrind}], we can also relate the \bg coefficients between the various frames as
\begin{eqnarray}
\alpha_{(20)}(k,m) &=& \int dp \left( \alpha_{(10)}(k,p)\alpha_{(21)}(p,m) + \beta_{(10)}(k,p)\beta^*_{(21)}(p,m) \right) \\
\beta_{(20)}(k,m) &=& \int dp \left( \alpha_{(10)}(k,p)\beta_{(21)}(p,m) + \beta_{(10)}(k,p)\alpha^*_{(21)}(p,m) \right)
\label{finalbeta}
\end{eqnarray}
The required expectation value of $N= a^\dagger_{(2)k}a_{(2)k}$ in the Minkowski vacuum $|0_M \rangle$ is then found out to be
\begin{eqnarray}
{\cal N}_{21} &=& \langle 0_M | a^\dagger_{(2)k}a_{(2)k} |0_M \rangle =\int d k^\prime |\beta_{(21)}({k^\prime, k})|^2 \nonumber \\ 
&=& \int dp \left[ |\beta_{(21)}(p,m)|^2 \frac{e^{\frac{\pi \omega_p}{g}}}{\sinh{\frac{\pi \omega_p}{g}}} + |\alpha_{(21)}(p,m)|^2 \frac{e^{-\frac{\pi \omega_p}{g}}}{\sinh{\frac{\pi \omega_p}{g}}}\right]
\label{directbetasquare}
\end{eqnarray}
As a consistency check note that in the limit $g^\prime \rightarrow 0$, that is, when $\beta_{(21)}(p,m) = 0$ and $\alpha_{(21)}(p,m) = \delta(p-m)$], we get back the Planckian spectrum corresponding to the Unruh bath at temperature $T = g/2\pi$ and similarly in the other limit $g \rightarrow 0$ we get the Unruh bath at temperature $T = g^\prime/2\pi$ .

It is also possible to derive the right hand side of Eq.[\ref{directbetasquare}], by the using the known properties of Rindler operators as follows. We want to evaluate
\begin{equation}
\langle n_{(2)q} \rangle = \langle 0_M| a^\dagger_{(2)q}a_{(2)q} |0_M\rangle
\label{nop}
\end{equation}
To find this expectation value, we first express $a^\dagger_{(2)q}$, $a_{(2)q}$ in terms of the Rindler's creation, annihilation operators $a^\dagger_{(1)q}$, $a_{(1)q}$ as
\begin{equation}
 a_{(2)q} = \int dp \left[ a_{(1)p}\alpha_{(21)}(p,q) + a^\dagger_{(1)p}\beta^*_{(21)}(p,q) \right]
\label{nop2}
\end{equation}
and then use 
\begin{equation}
 \langle 0_M| a^\dagger_{(1)p}a_{(1)p} |0_M\rangle = \delta(0) \left[ \frac{1}{e^{\frac{2\pi \omega_p}{g}} -1} \right]
\label{nop3}
\end{equation}
Using Eqn.[\ref{nop2}] in Eqn.[\ref{nop}], we get
\begin{eqnarray}
\langle n_{(20)} \rangle &=& \langle 0_M|\int \int dp dp^\prime \biggl[ \alpha^*_{(21)}(p^\prime,q)\alpha_{(21)}(p,q)[a^\dagger_{(1)p^\prime}a_{(1)p}] \nonumber \\
&& + \beta_{(21)}(p^\prime,q)\beta^*_{(21)}(p,q)[a_{(1)p^\prime}a^\dagger_{(1)p}] \biggr] |0_M\rangle \nonumber \\
& = &\int dp \left[ |\beta_{(21)}(p,q)|^2 \frac{e^{\frac{\pi \omega_p}{g}}}{\sinh{\frac{\pi \omega_p}{g}}} + |\alpha_{(21)}(p,q)|^2 \frac{e^{-\frac{\pi \omega_p}{g}}}{\sinh{\frac{\pi \omega_p}{g}}}\right] 
\end{eqnarray}
which is essentially same as the right hand side of Eqn.[\ref{directbetasquare}]. 
The above equation can be further simplified to get
\begin{eqnarray}
\langle n_{(20)} \rangle &=& \frac{1}{g E} \int \frac{d \omega_k}{2\pi\omega_k}\left[ \frac{1}{  \left( 1 - e^{-\beta \omega_k} \right)\left( e^{\frac{E 2 \pi}{g}} - 1 \right)}+ \frac{1}{  \left( e^{\beta \omega_k} - 1\right)\left( 1- e^{\frac{-E 2 \pi}{g}} \right)} \right] \nonumber \\
\label{thermalbathN}
\end{eqnarray}
Let us next define $n_k^{\beta} = 1/(e^{\beta \omega_k} -1)$ and $n_p^{\beta^\prime} = 1/(e^{\beta^\prime \omega_p} -1)$. In terms of these variables, the integrand in Eq.[\ref{thermalbathN}] can be written as
\begin{eqnarray}
I &=& n_k^{\beta} +n_p^{\beta^\prime} + 2n_p^{\beta^\prime} n_k^{\beta}
\end{eqnarray}
The above form is similar to that obtained in the reduced density matrix formalism for a uniformly accelerated observer moving in the thermal \textit{state} of the quantum field and can be shown to have an interpretation in terms of the spontaneous and stimulated emission of Rindler particles \cite{san2}.

\section{Detector response}

We next want  to compare (i) the spectrum of particles detected by a suitable observer in the Rindler-Rindler spacetime in the Minkowski vacuum with (ii) that detected by an observer accelerating uniformly through  a real thermal bath. For this purpose, we consider an Unruh-Dewitt detector \cite{unruhdet} moving on an uniformly accelerated trajectory in an inertial thermal bath and calculate its excitation rate. Such a calculation has been attempted before (see for e.g., \cite{accthermalbath, accthermalbath2}). Here, we shall demonstrate the same in a manner suitable for our present purpose.

The detector essentially consists of a two level quantum system linearly coupled to the quantum field. We analyze its excitation probability when it is in motion; particularly when it is moving on an uniformly accelerated trajectory. Let the interaction of the detector with the scalar field be given by a monopole coupling of the form $c m(\tau) \phi(\tau)$ where $c$ is a small coupling constant and $m$ is the detector's monopole moment. It is well known that using linear perturbation theory the excitation rate for the detector is given by
\begin{eqnarray}
 R(E) = \int^{\infty}_{-\infty} d\triangle \tau \; e^{-i E \triangle \tau} G^+(\tau - \tau^\prime )
 \label{detresponse}
\end{eqnarray}
where $G^+(\tau - \tau^\prime )$ is the Wightman function defined as 
\begin{eqnarray}
G^+(\tau - \tau^\prime ) = \langle \psi| \phi(X(\tau))\phi(X(\tau^\prime))| \psi \rangle
\label{whitemandef}
\end{eqnarray}
where the state $| \psi \rangle$ is the initial state of the scalar field at $\tau \rightarrow -\infty$. In the present context, we are interested in the case when the initial \textit{state} of the field corresponds to  a thermal state with a temperature $T$. For this purpose, we define a nonzero temperature Green function, also called as the thermal Green's function, constructed by replacing the vacuum expectation values by an thermally weighted  ensemble average as follows  
\begin{equation}
 G^+_\beta(t,\textbf{x};t^\prime, \textbf{x}^\prime ) = tr \left[ e^{-\beta H} \phi(x)\phi(x^\prime) \right]/tr \left[ e^{-\beta H} \right]
\end{equation}
Here, the thermal nature of the field  is essentially captured by the density matrix $\rho = e^{-\beta H}$ with $\beta^{-1} $ being the temperature and $H$ the Hamiltonian of the scalar field. It can be shown that the above expression in $n+1$ dimensional flat spacetime  takes the form
\begin{eqnarray}
 G^+_\beta(T,{\bf X};T^\prime, {\bf X}^\prime ) = \frac{1}{(2\pi)^n} \int d^n {\bf k} \left[ \frac{e^{-i\omega_{{\bf k}} (T - T^\prime) +i {\bf k} \cdot ({\bf X} - {\bf X}^\prime) }}{2\omega_{{\bf k}} \left( 1 - e^{-\beta \omega_{{\bf k}}}\right)} -  \frac{e^{i\omega_{{\bf k}} (T - T^\prime) -i {\bf k} \cdot ({\bf X} - {\bf X}^\prime) }}{2\omega_{{\bf k}} \left( 1 - e^{\beta \omega_{{\bf k}}}\right)} \right]
\label{thermalndim}
\end{eqnarray}
By analogy with the expression for the response rate of the Unruh De-Witt detector of Eq.[\ref{detresponse}], one can similarly define a response rate of a detector moving in an inertial thermal bath using the thermal Green's function as 
\begin{eqnarray}
 R(E) = \int^{\infty}_{-\infty} d\triangle \tau \; e^{-i E \triangle \tau} G^+_\beta(\tau - \tau^\prime )
 \label{detresponsethermal}
\end{eqnarray}
Here, one caveat to be noted is that the above expression is not a first principle derivation, starting from linear perturbation theory, but rather an extension of the formula in Eq.[\ref{detresponse}]. However, in the current context, it is instructive to note the following: if one considers the initial state of the scalar field to be described by a Fock state $| \Psi \rangle$ such that (i) it is an eigenstate of the Hamiltonian $H =\sum_k \omega_k a^\dagger_{{\bf k}} a_{{\bf k}}$ and (ii) it satisfies 
\begin{equation}
 a^\dagger_{{\bf k}} a_{{\bf k}} | \Psi \rangle = \frac{1}{e^{\beta \omega_{{\bf k}}} -1} | \Psi \rangle
\label{statedef}
\end{equation}
then it turns out that 
the Wightman function as defined in Eq.[\ref{whitemandef}] for the Fock state $| \Psi \rangle$ takes the following form 
\begin{eqnarray}
 G^+_{\Psi}(T,{\bf X};T^\prime, {\bf X}^\prime ) = \frac{1}{(2\pi)^n} \int d^n {\bf k} \left[ \frac{e^{-i\omega_{{\bf k}} (T - T^\prime) +i {\bf k} \cdot ({\bf X} - {\bf X}^\prime) }}{2\omega_{{\bf k}} \left( 1 - e^{-\beta \omega_{{\bf k}}}\right)} -  \frac{e^{i\omega_{{\bf k}} (T - T^\prime) -i {\bf k} \cdot ({\bf X} - {\bf X}^\prime) }}{2\omega_{{\bf k}} \left( 1 - e^{\beta \omega_{{\bf k}}}\right)} \right]
\label{whitemanndim}
\end{eqnarray}
Comparing Eq.[\ref{thermalndim}] and eq.[\ref{whitemanndim}], we can see that they are identical. Hence, to the lowest order in the field fluctuations, the state $| \Psi \rangle$ defined by Eq.[\ref{statedef}] acts as a thermal state. However, as is obvious, higher moments such as $~ \langle \phi(x)^3 \rangle$,$~ \langle \phi(x)^4 \rangle$, etc will differ. Therefore, all physical quantities which depend only on the smallest moment of the field fluctuations will lead to the same result whether one consider the thermal Green's function or the Wightman function constructed from the Fock state. The response of the Unruh-Dewitt detector coupled to the Fock state $| \Psi \rangle$ is one such physical quantity. We therefore have  
\begin{eqnarray}
 R(E) & = & \int^{\infty}_{-\infty} d\triangle \tau \; e^{-i E \triangle \tau} G^+_\Psi(\tau - \tau^\prime ) \nonumber \\
&=& \int^{\infty}_{-\infty} d\triangle \tau \; e^{-i E \triangle \tau} G^+_\beta(\tau - \tau^\prime )
\end{eqnarray}
One now has a sufficient motivation to start with the expression Eq.[\ref{detresponsethermal}]. We now proceed to calculate the response for the detector moving on an accelerated trajectory 
\begin{eqnarray}
T=\frac{1}{g}\sinh{g \tau}  \; , \; \; \;\;\;\;\;  X = \frac{1}{g}\cosh{g \tau}
\end{eqnarray}
Substituting the above equation in Eq.[\ref{detresponsethermal}] and performing both the time integrals, we get the response function to be 
\begin{eqnarray}
 {\cal R}(E) &=& \int \frac{dk}{2\pi} \frac{1}{2 \omega_k \left( 1 - e^{-\beta \omega_k} \right)} \left| \int e^{-i E \tau} e^{-i \omega_k (T-X)} d\tau \right|^2 \nonumber \\
&& - \int \frac{dk}{2\pi} \frac{1}{2 \omega_k \left( 1 - e^{\beta \omega_k} \right)} \left| \int e^{-i E \tau} e^{i \omega_k (T-X)} d\tau \right|^2 \\
&=&\frac{1}{g E} \int \frac{d \omega_k}{2\pi\omega_k}\left[ \frac{1}{  \left( 1 - e^{-\beta \omega_k} \right)\left( e^{\frac{E 2 \pi}{g}} - 1 \right)}+ \frac{1}{  \left( e^{\beta \omega_k} - 1\right)\left( 1- e^{\frac{-E 2 \pi}{g}} \right)} \right] \nonumber \\
\label{detresponsefinal}
\end{eqnarray}
The above expression is same as the expression obtained in Eq.[\ref{thermalbathN}].

\section{Conclusions}

In this paper, we have investigated the issue whether one can exploit the thermality of the Davies-Unruh bath and make it act as a proxy for  a genuine thermal bath. In particular, we wanted the explore whether the effects that one encounters in the study of quantum fields in a real thermal bath from the perspective of the uniformly accelerated observer can be mimicked by the utilizing the Davies-Unruh bath as a proxy thermal bath for a special set of observers. We found that this is indeed the case.

We demonstrated that for the set of observers described by the solutions of Eq.[\ref{diffeqn}], the Davies-Unruh bath does appear to be populated in a manner identical  to that observed by an observer uniformly accelerating through  a genuine thermal bath. To determine these set of trajectories, we calibrated the particle content at a particular instant $T = 0$, such that the vacuum state of the Rindler observer appears thermal to the Rindler-Rindler observers. Then,  we compared the particle content in the Davies-Unruh bath seen by the  Rindler-Rindler observer to that obtained by the response of an uniformly accelerated detector moving in a genuine thermal bath. We found both the results to be identical. We interpret this similarity as indicating further evidence of the indistinguishablity between thermal and quantum fluctuations along the lines discussed in ref. \cite{san}.
The geometrical properties of Rindler-Rindler spacetime, the natural observers in this spacetime and the extension of these results to (1+3) dimensions are worthy of further investigation.
\section*{Acknowledgements}
The research of TP is partially supported by the J.C. Bose fellowship of DST, India

\end{document}